\documentclass[aps,prd,groupedaddress]{revtex4}
\def\ga{\mathrel{\raise.3ex\hbox{$>$\kern-.75em\lower1ex\hbox{$\sim$}}}}
\def\la{\mathrel{\raise.3ex\hbox{$<$\kern-.75em\lower1ex\hbox{$\sim$}}}}

\def\lsim{\mathrel{\rlap{\lower4pt\hbox{\hskip1pt$\sim$}}
    \raise1pt\hbox{$<$}}}                
\def\gsim{\mathrel{\rlap{\lower4pt\hbox{\hskip1pt$\sim$}}
    \raise1pt\hbox{$>$}}}
\usepackage{graphicx,color}

\usepackage{amsmath}
\usepackage{amssymb}
\usepackage{bbm}
\usepackage{epsfig}
\begin{document}
\title{Search for a light fermiophobic Higgs boson produced via
gluon fusion at Hadron Colliders}
\author{Abdesslam Arhrib$^{1,3}$~\footnote{aarhrib@ictp.it},
Rachid Benbrik$^{2,3}$~\footnote{rbenbrik@phys.cycu.edu.tw}, R.B.
Guedes$^{4}$~\footnote{renato@cii.fc.ul.pt} and R.
Santos$^{4,5}$~\footnote{rsantos@cii.fc.ul.pt}}
\affiliation{$^{1}$D\'epartement de Math\'ematiques, Facult\'e des
Sciences et
Techniques, B.P 416 Tangier, Morocco\\
$^{2}$ Chung Yuan Christian University, Department of Physics,
Chung-Li, Taiwan 320, R.O.C.\\
$^{3}$ LPHEA, D\'epartement de Physique,Facult\'e des Sciences-Semlalia
B.P 2390 Marrakech, Morocco\\
$^{4}$ Centro de F\'{\i}sica Te\'orica e Computacional, Faculdade
de Ci\^encias, Universidade de Lisboa, Avenida Professor Gama
Pinto, 2, 1649-003
Lisboa, Portugal \\
$^{5}$ Department of Physics, Royal Holloway, University of
London, Egham, Surrey TW20 0EX United Kingdom}

\date{\today}

\begin{abstract}
\noindent In this study, we propose new Higgs production
mechanisms with multi-photon final states in the fermiophobic
limit of the Two Higgs Doublet Model. The processes are: $gg\to
hh$, $gg\to H h$ followed by $H \to h h$ and $gg\to A h$ followed
by $A \to h Z$. In the fermiophobic limit, $gg\to h h$ and $gg\to
A h \to  h h Z$ would give rise to 4$\gamma$ signature while
$gg\to H h \to h h h$ can give a 6$\gamma$ final state. We show
that both the Fermilab Tevatron and CERN's Large Hadron Collider
can probe a substantial slice of the parameter space in this
fermiophobic scenario of the Two Higgs Doublet Model. If observed
the above processes can give some information on the triple Higgs
couplings involved.
\end{abstract}

\maketitle

\section{Introduction}
\noindent
There are several extensions of the Standard Model (SM) with an
enlarged scalar sector. Some of these extensions allow for Higgs
with reduced or even no couplings to the fermions. They are
referred to as fermiophobic Higgs scenarios in the literature
\cite{fermio}. The D0 collaboration has recently presented new
results on fermiophobic Higgs searches \cite{D03gamma,2008it}. In
\cite{D03gamma} they have searched for a fermiophobic Higgs
produced in association with a charged Higgs. The full process is
$p \bar{p} \rightarrow h H^{\pm} \rightarrow hh W^{\pm \, *}
\rightarrow 4 \gamma + X$ and  was proposed in
\cite{Akeroyd:2003bt,Akeroyd:2005pr,Akeroyd:2003xi}. D0 required
at least three photons in the final state for maximizing signal
efficiency. For each pair of values of $(\tan \beta, m_{H^{\pm}})$
a bound on the fermiophobic Higgs mass was set. In \cite{2008it},
D0 has performed a search for the inclusive production of
di-photon final states via the Higgsstrahlung and vector boson
fusion processes: $p \bar{p} \rightarrow h V \rightarrow \gamma
\gamma + X$ and $p \bar{p} \rightarrow V V \rightarrow h
\rightarrow \gamma \gamma + X$, respectively, with a total
integrated luminosity of 1.10 $\pm$ 0.07 $fb^{-1}$. A lower $m_h$
bound of 100 GeV was obtained in a benchmark scenario that assumes
hVV (V= W, Z) couplings to be exactly the same as in the SM and
all fermion branching ratios to be exactly zero.
\\
All LEP collaborations have searched for a fermiophobic Higgs in
the channel $e^+ e^- \rightarrow h (\rightarrow \gamma \gamma) Z$
~\cite{Abbiendi:2002yc,Abreu:2001ib,Heister:2002ub,Achard:2002jh}.
The combination of all results~\cite{Rosca:2002me} yielded the
lower bound for the fermiophobic Higgs mass of 109.7 GeV, at 95\%
confidence level, which again is valid only in the above benchmark
scenario. Searches in the $e^+ e^- \rightarrow h A$ channel were
also performed at LEP with lower bounds derived for $m_h+ m_A$
(see references ~\cite{Abbiendi:2002yc,Abreu:2001ib} for details).
The new D0 bound~\cite{2008it} on $m_h$ is weaker but a larger
region of the model's parameter space is covered. The channel $qq'
\rightarrow V^* \rightarrow h (\rightarrow \gamma \gamma) \, V$
had already been used at the Tevatron by the
CDF~\cite{Abbott:1998vv} and by the D0~\cite{Affolder:2001hx}
collaborations to set limits of 78.5 GeV and 82 GeV at 95\%
confidence level, respectively, on the fermiophobic Higgs mass. It
is expected~\cite{Mrenna:2000qh,Abazov:2005un,Landsberg:2000ht}
that all Tevatron bounds will improve once the data collected at
$2 fb^{-1}$ luminosity is analyzed.
\\
There are however other ways of producing a four photon final
state in a fermiophobic scenario. In this letter we will consider
all fermiophobic Higgs production processes with at least three
photons in the final state produced via gluon fusion. As shown in
\cite{D03gamma}, a signal with at least three photons is very easy
to extract at the Tevatron. The most relevant process for the
analysis is $gg \rightarrow hh \rightarrow 4\, \gamma$ but we can
also have $gg \rightarrow hH \rightarrow hhh \rightarrow 6\,
\gamma$ and $gg \rightarrow hA \rightarrow hhZ \rightarrow 4\,
\gamma + X$. We will show that both the Tevatron and the Large
Hadron Collider (LHC) can probe a substantial region of the
parameter space. We will discuss as well the complementarity
between the different production modes.
\\
This paper is structured as follows: we will review the
fermiophobic model in section II and then proceed to look at the
all available theoretical and experimental bounds in section III.
In section IV, we will discuss in detail the production process
and in section V the signal. Analysis of the results and
conclusions will be presented in section VI.

\section{The Fermiophobic THDM}
\noindent To define our notation we start with a brief review of
the two-Higgs doublet potential used here. The potential chosen is
the most general, renormalizable, CP-conserving potential,
invariant under $SU(2) \otimes U(1)$ that one can build with two
complex Higgs doublets. It can be written as
\begin{eqnarray}
 V(\Phi_{1}, \Phi_{2})& & =  \lambda_{1} ( |\Phi_{1}|^2-v_{1}^2)^2
+\lambda_{2} (|\Phi_{2}|^2-v_{2}^2)^2+
\lambda_{3}((|\Phi_{1}|^2-v_{1}^2)+(|\Phi_{2}|^2-v_{2}^2))^2
+\nonumber\\ [0.2cm]
&  & \lambda_{4}(|\Phi_{1}|^2 |\Phi_{2}|^2 - |\Phi_{1}^+\Phi_{2}|^2  )+
\lambda_{5} [\Re e(\Phi^+_{1}\Phi_{2})
-v_{1}v_{2}]^2+ \lambda_{6} [\Im m(\Phi^+_{1}\Phi_{2})]^2
\label{higgspot}
\end{eqnarray}
where $\Phi_1$ and $\Phi_2$ have weak hypercharge Y=1, $v_1$ and
$v_2$ are respectively the vacuum expectation values of $\Phi_1$
and $\Phi_2$ and the $\lambda_i$ are real--valued parameters. Note
that this potential violates the discrete symmetry $\Phi_i\to
-\Phi_i$ only softly  by the dimension two term $\lambda_5 \Re
e(\Phi^+_{1}\Phi_{2})$. The hard breaking terms (dimension four)
of the discrete symmetry have been set to zero.
As in all other THDM, we end up with two CP-even Higgs states
usually denoted by $h$ and $H$, one CP-odd state, $A$ and two
charged Higgs bosons, $H^{\pm}$. The potential in
eq.~(\ref{higgspot}) has 8 parameters (including $v_1$ and $v_2$).
The combination $v^2=v_1^2 + v_2^2$ is fixed as usual by the
electroweak breaking scale through $v^2=(2\sqrt{2} G_F)^{-1}$.  We
are thus left with 7 independent parameters; namely
$(\lambda_i)_{i=1,\ldots,6}$, and $\tan\beta \equiv v_2/v_1$.
Equivalently, we can take instead 
\begin{eqnarray} m_{h}\quad ,
\quad m_{H} \quad , \quad m_{A} \quad , \quad m_{H^\pm} \quad ,
\quad \tan\beta \quad , \quad \alpha \quad \rm{and} \quad
\lambda_5. \label{parameters}
\end{eqnarray}
as the 7 independent parameters.
The angle $\beta$ is the rotation angle from the group eigenstates
to mass eigenstates in the CP-odd and charged sector. The angle
$\alpha$ is the corresponding rotation angle for the CP-even
sector.\\
In a general THDM it is possible to couple just one doublet to all
fermions by choosing an appropriate symmetry for both the fermions
and the scalars. This model is known as THDM type I in the
literature. Like in the SM, where just one doublet couples to all
fermions, each scalar couples to the different fermions with the
same coupling constant. However, unlike the SM, the couplings are
now proportional to the rotation angles $\alpha$ and $\beta$. For
instance, the lightest CP-even Higgs couples to the fermions as
$\cos \alpha/\sin \beta \, \, g_{h \bar{f} f}^{SM}$.
By choosing $\cos \alpha = 0$, the lightest CP-even Higgs
decouples from all fermions. It is usually referred to as a
fermiophobic Higgs scalar~\cite{fermio}. This way the heavy
CP-even scalar will acquire larger couplings to the fermions than
the corresponding SM couplings. The remaining scalars are not
affected by this choice as they do not couple proportionally to
$\alpha$.
\begin{figure}[htbp]
  \begin{center}
    \epsfig{file=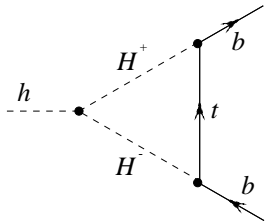,width=3.5cm}
    \caption{Feynman diagram of the largest contribution to $h^0\rightarrow b \bar{b}$}
    \label{fig:fig_hff}
  \end{center}
\end{figure}
However,
$h$ can still decay to two fermion pairs via \mbox{$h \rightarrow
  W^{*}W (Z^{*}Z) \rightarrow 2\, \bar{f} f$} or \mbox{$h
  \rightarrow W^* W^* (Z^* Z^*) \rightarrow 2\,\bar{f} f$}. We will
include these decays in our analysis. It is worth pointing out
that these processes occur near the $W (Z)$ threshold.  Decays of
$h$ to two fermions can also be induced by scalar and gauge boson
loops (see e.g. fig.~\ref{fig:fig_hff}). In the THDM, the angle
$\alpha$ has to be renormalized to render $h \rightarrow f\bar{f}$
finite. However, at $\alpha=\pi/2$, all one-loop decays $h
\rightarrow f\bar{f}$ are finite. Thus we can impose the following
condition for the renormalization constant $\delta\alpha$: the
renormalized one-loop decay width for $h \rightarrow f\bar{f}$ is
equal to the finite unrenormalized decay width. This condition is
equivalent to setting $[\delta\alpha]_{\alpha=\pi/2}=0$. In
\cite{Brucher:1999tx} we have checked that this condition holds
for all fermions. The only relevant one-loop decay is $h
\rightarrow b\bar{b}$ due to a large contribution of the Feynman
diagram shown in fig.~\ref{fig:fig_hff} to the total decay
width~\footnote{The coupling $[H^+\bar{t}b]$ is proportional to
the
  $t$-quark mass.}.  Thus, on one hand, $h$ is not completely
fermiophobic at $\alpha=\pi/2$, and on the other hand, all decays
$h \rightarrow f\bar{f}$ but $h \rightarrow b\bar{b}$ are almost
zero even at one-loop level. Nevertheless it is possible, although
hard, to find regions of the parameter space where $h \rightarrow
b\bar{b}$ has a sizeable effect in the fermiophobic Higgs
signature. With this in mind it is obvious that we are working
with a version of the model with one less parameter than the
general CP-conserving THDM.

\section{Theoretical and experimental bounds}
\noindent
\label{sec:bounds} In our analysis we take into account the
following experimental and theoretical constraints. Note however
that we will use the theoretical constraints as guides and explain
in each case how they would affect the results given in the plots.
We believe that if Nature did not provide a fermiophobic Higgs it
should be disallowed ultimately by experiment.

\begin{itemize}

\item Experimental bounds from LEP; the LEP collaborations have
set bounds on the mass of a fermiophobic Higgs as a function of
$\tan \beta$. The most stringent bound comes from the combination
of all LEP experiments given in~\cite{Rosca:2002me}. The 95\% CL
limit on $\tan \beta$ is about 18 for $m_h=20$ GeV. From $m_h=20$
GeV until $m_h=70$ GeV, the bound oscillates about $\tan \beta
=10$. For $m_h > 70$ GeV, $\tan \beta > 10$ is already a
conservative bound. In the plots $\tan \beta = 10$ is used because
much higher values of $\tan \beta$ would violate perturbativity
constraints. Note however that as we will show, $gg \to hh$ is
independent of $\tan \beta$ and for $\tan \beta
> 10$ the dependence of all other relevant processes is
negligible.
\\
There is another bound on the fermiophobic Higgs coming from $hA$
production at LEP which constrains severely the value of $\tan
\beta$ especially for small $h$ masses. If however we take the $A$
mass to be above 150 GeV, the $hA$ production is no longer a
constraint for all values of $\tan \beta$.
\\
Finally, if $\tan \beta$ is large, the $ZZh$ coupling is
suppressed while the non-fermiophobic CP-even Higgs $H$ will
couple to the $Z$ bosons with almost the SM strength. Therefore,
in the plots shown, the minimum value of the $H$ mass is 100 GeV
and most cases presented are for an $H$ mass above 120 GeV.
\\
\item As already stated in the introduction, the D0 and CDF
experiments recently reported searches for a fermiophobic Higgs in
two different channels. D0 searched~\cite{2008it} for the
inclusive production of di-photon final states via the
Higgsstrahlung and vector boson fusion. The bound with the model
benchmark described in the introduction is weaker than the LEP
bound but it spans a larger region of the parameter space. The
search~\cite{D03gamma} in the $p \bar{p} \rightarrow h H^{\pm}
\rightarrow hh W^{\pm \, *} \rightarrow 4 \gamma + X$ channel sets
a bound on the fermiophobic Higgs mass for each pair of values of
$(\tan \beta, m_{H^{\pm}})$. We will take all these bounds into
consideration in our analysis.

\item The extra contributions to the $\delta\rho$ parameter from
the Higgs scalars \cite{Rhoparam} should not exceed the current
limits from precision measurements \cite{pdg4}: $ |\delta\rho| \la
10^{-3}$. Such an extra contribution to $\delta\rho$ vanishes in
the limit $m_{H^\pm}=m_{A}$. To ensure that  $\delta\rho$ will be
within the allowed range whenever possible we allow only a small
splitting between $m_{H^\pm}$ and $m_{A}$.

\item Recently, it has been shown in Ref.~\cite{Oslandk} that for
THDM models of the type II, data on $B\to X_s \gamma$ imposes a
lower limit of $m_{H^\pm} \ga 290$\,GeV. In THDM type I, there is
no such  constraint on the charged Higgs mass. Therefore, in our
numerical analysis which is valid for THDM type I we will ignore
the limit on the charged Higgs.


\item The scalar sector can also be constrained using
perturbativity constraints on $\lambda_i$ \cite{unit1,abdesunit}.
In the present study we will not impose those constraints in order
to quantify the optimal cross sections and scan over all
parameters space. In fact, in the fermiophobic limit, the process
$$gg\to hh$$ for example depends only on $m_h$, $m_H$ and
$\lambda_5$. As we will explain later, the $\tan\beta$ dependence
in  $gg\to hh$ drops out. As a result, for a given $m_h$, $m_H$
and $\lambda_5$, one can tune $\tan\beta$, $m_A$ and $m_{H\pm}$ in
order to satisfy perturbativity constraints.



\item From the requirement of perturbativity for the top and
bottom Yukawa couplings \cite{berger}, $\tan\beta$ is constrained
to lie in the range $0.3\leq \tan\beta \leq 100$. But it turns out
that from perturbativity argument on $\lambda_i$, moderate values
of $\tan\beta$ less than about 10 are preferred.

\end{itemize}
In order to respect perturbativity constraints we will use
moderate values for $M_h$, $M_H$, $M_A$ and $M_{H^\pm}$.

\section{Production cross sections}
\label{sec:cross}
\begin{figure}[htbp]
  \begin{center}
    \epsfig{file=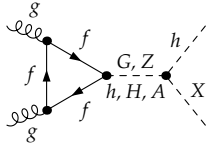,width=4cm}
    \caption{Triangle contribution to $gg \rightarrow hX$ with $X=H,h,A$.}
    \label{fig:fig_verthhgg}
  \end{center}
\end{figure}

\noindent
The process $pp (\bar{p}) \rightarrow hh$ has both tree level
contributions mediated by Higgs exchange from $q \bar{q}
\rightarrow H^*,h^* \rightarrow hh$ and one loop contributions
from gluon fusion $gg\rightarrow hh$. The tree level contribution
is proportional to the quark masses which will be neglected for
the Tevatron energies. We have also checked that even in the large
$\tan \beta$ limit the production process $q \bar{q} \rightarrow
hh$ is also negligible for the LHC when compared to $g g
\rightarrow hh$.
The process $gg \rightarrow hh$ occurs only at the one-loop level.
As we will see, even if loop suppressed, this process can still be
enhanced by the strong QCD coupling as well as by the heavy Higgs
$H$ resonant effect when it can decay to two light CP even $h$
scalars. There are two types of diagrams that participate in the
process $gg \rightarrow hh$.
\begin{figure}[!h]
  \begin{center}
    \epsfig{file=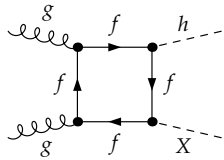,width=4cm}
    \caption{Box contribution to $gg \rightarrow hX$ with $X=H,h,A$ and $f$ is a fermion.
    Just one topology is shown.}
    \label{fig:fig_boxhhgg}
  \end{center}
\end{figure}
The box diagram, as shown in fig~\ref{fig:fig_boxhhgg} represents
just a generic contribution to the process. We have also included
all other quarks and because both the initial and the final state
have identical particles, there is a total of six diagrams for
each flavor.
The second type are the vertex diagrams which again are six for
each flavor. In fig.~\ref{fig:fig_verthhgg} we show just the
representative diagrams with a generic fermion in the loop. In the
fermiophobic limit, the top loop is always the dominant
contribution.
The other two processes, $pp (\bar{p}) \to H \, h$ and $pp
(\bar{p}) \to A \, h$ have a similar structure (vertex and box) to
$pp\to h h$, except for one additional contribution to the vertex
diagrams in $pp  (\bar{p}) \to A \, h$
which is the $Z$ boson and the Goldstone boson s-channel exchange.\\
However, in the fermiophobic limit, all box contributions vanish
and the same is true for all vertex contribution with s-channel
Higgs ($h$) exchange due to the fact that the fermiophobic Higgs
coupling to fermions is zero. Since the box contributions drop out
in the fermiophobic limit, the two processes $pp (\bar{p}) \to h
h, H h$ are directly proportional to the pure scalar couplings $H
h h$ and $H H h$. The third process $pp (\bar{p}) \to A h$ is
sensitive both to pure scalar couplings $A A h$ and $A G h$ as
well as to the gauge coupling $Z A h$. Hereafter we list the pure
scalar coupling in the fermiophobic limit (FL) needed for our
study:
\begin{eqnarray}
\lambda_{H h h}^{FL} &=&
-\frac{e \lambda_5 v^2 \, \sin \beta}{4 m_W s_W }\propto \lambda_5\frac{\tan\beta}{\sqrt{1+\tan^2\beta}}  \label{hll} \\
\lambda_{H H h}^{FL} &=&
\frac{e \lambda_5 v^2 \, \cos \beta }{4 m_W s_W}\propto \lambda_5\frac{1}{\sqrt{1+\tan^2\beta}} \label{hhl}\\
\lambda_{A A h}^{FL} &=& \frac{-e}{4 m_W s_W \,  \cos
\beta}\bigg[-2 \sin^2 \beta \, m^2_{h} +
 \lambda_5 v^2  - 4 \cos^2 \beta  \, m^2_{A} \bigg]
\label{aal}\\
\lambda_{A G h}^{FL} &=& \frac{e \,  \sin \beta}{2 m_W
s_W}\bigg( m^2_{A} - m^2_{h}\bigg)\label{agl}\\
\lambda_{hH^\pm H^\mp}^{FL}&=&
\frac{e}{2 m_W s_W\sin 2\beta}\bigg[2 \sin^3\beta
m^2_{h} -  \lambda_5 v^2 \sin\beta- 2 \sin{2\beta} \cos{\beta}
m^2_{H^\pm} \bigg]\label{lhphm}
\end{eqnarray}
Note that the $H h h$ and $H H h$ couplings are directly
proportional to $\lambda_5$ while the $A A h$ coupling depend both
on $\lambda_5$, $m_{A}$ as well as on $m_{h}$. It is clear that in
the case of exact discrete $Z_2$ symmetry ($\lambda_5=0$), both $H
h h$ and $H H h$ would vanish. In the fermiophobic limit, all
fermions couple to each scalar with the same strength. The
$\bar{f} f H$ coupling is proportional to $1/\sin \beta$.
Therefore from eq.~\ref{hll} we conclude that the $\beta$ angle
dependence cancels out in the cross section, \textit{that is},
$\sigma_{gg \to hh}$ depends only on $m_h$, $m_H$ and $\lambda_5$.
The $\bar{f} f A$ coupling is proportional to $\cos \beta/\sin
\beta$ which means that in the large $\tan \beta$ limit the
$\beta$ angle dependence is also very mild except for very large
$h$ and/or $A$ masses. Those cases will not be included in our
study. Finally, note that not only $\sigma_{gg \to hh}$ vanishes
in the limit $\lambda_5=0$ but also $\sigma_{gg \to Ah}$, in the
high $\tan \beta$ limit becomes negligible when $\lambda_5= 0$ due
to to the smallness of the
values of the masses involved.\\
The one-loop amplitudes were generated and calculated with
 the packages FeynArts \cite{feynarts} and FormCalc \cite{formcalc}.
The scalar integrals were evaluated with  LoopTools
\cite{looptools}.
\subsubsection{Numerical results}
\noindent
In this section, we present our numerical results. As stated
earlier, to avoid the LEP bounds on the fermiophobic Higgs we will
fix $\tan\beta$ to be of the order 10 for the processes $gg \to
Ah$ and $gg \to Hh$. This way we suppress
the $ZZh$ coupling while keeping perturbativity bounds on $\lambda_i$
within the allowed range. \\
The first consequence of this choice of $\tan\beta\approx 10$ is
that the coupling $Hhh$ is enhanced ($\sin\beta\approx 1$) while
$HHh$ is suppressed (see eqs.~\ref{hll}, \ref{hhl}). As we have
discussed, in the fermiophobic limit, the process $pp\to hh$ has
only vertex contribution through s-channel heavy Higgs ($H$)
exchange. Therefore, the cross section for $pp (\bar{p}) \to hh$
depend both on $m_H$ as well as on $Hhh$ coupling which is
proportional to $\lambda_5$. There are two sources of enhancements
for $pp\to hh$. The first one is to take $Hhh$ (or equivalently
$\lambda_5$) as large as possible. The second one is when $hh$
production is resonant, that is, $m_{H}\approx 2m_{h}$.
\begin{figure}[t!]
\centering
\includegraphics[height=2.7in]{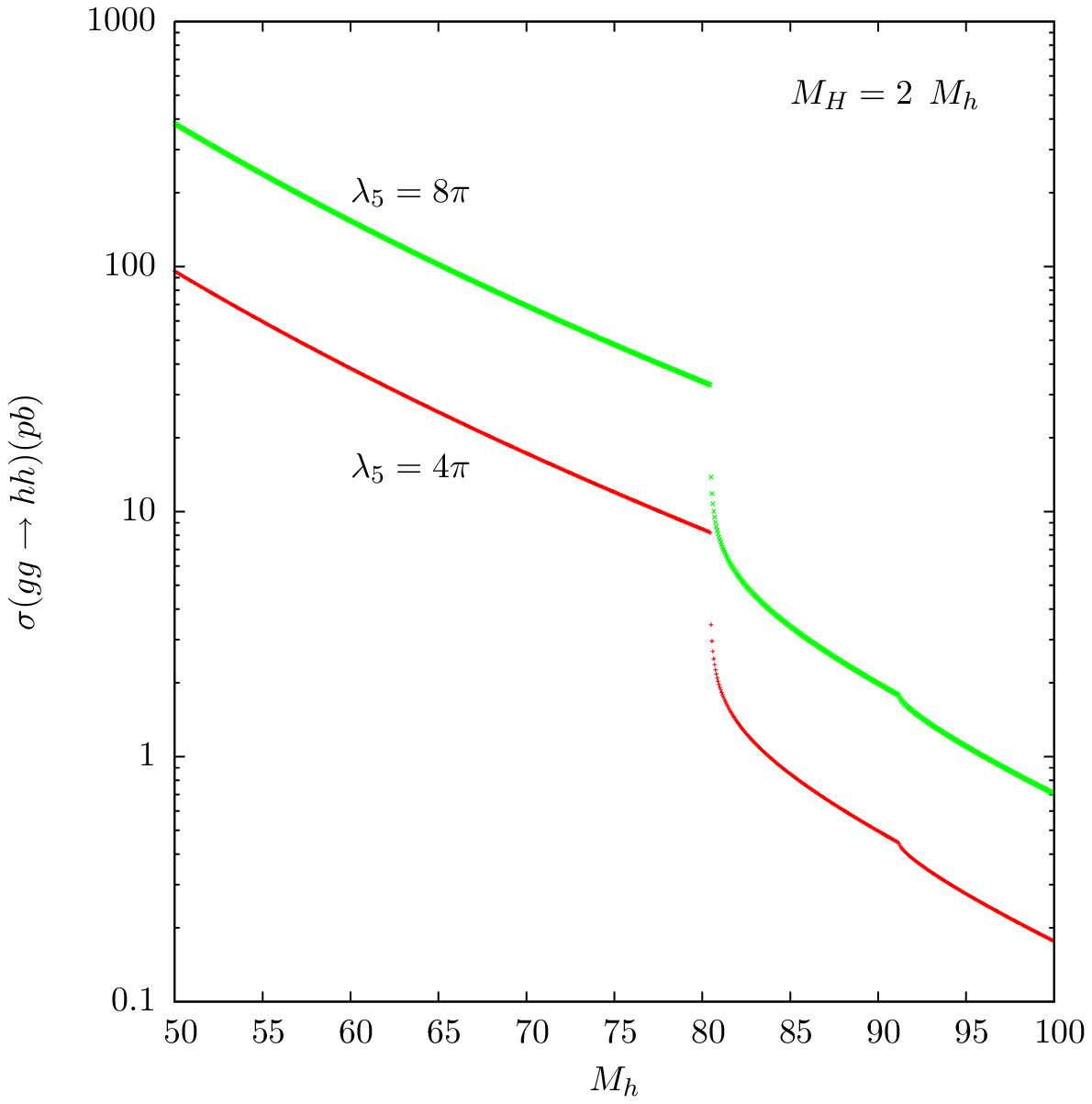}
\includegraphics[height=2.7in]{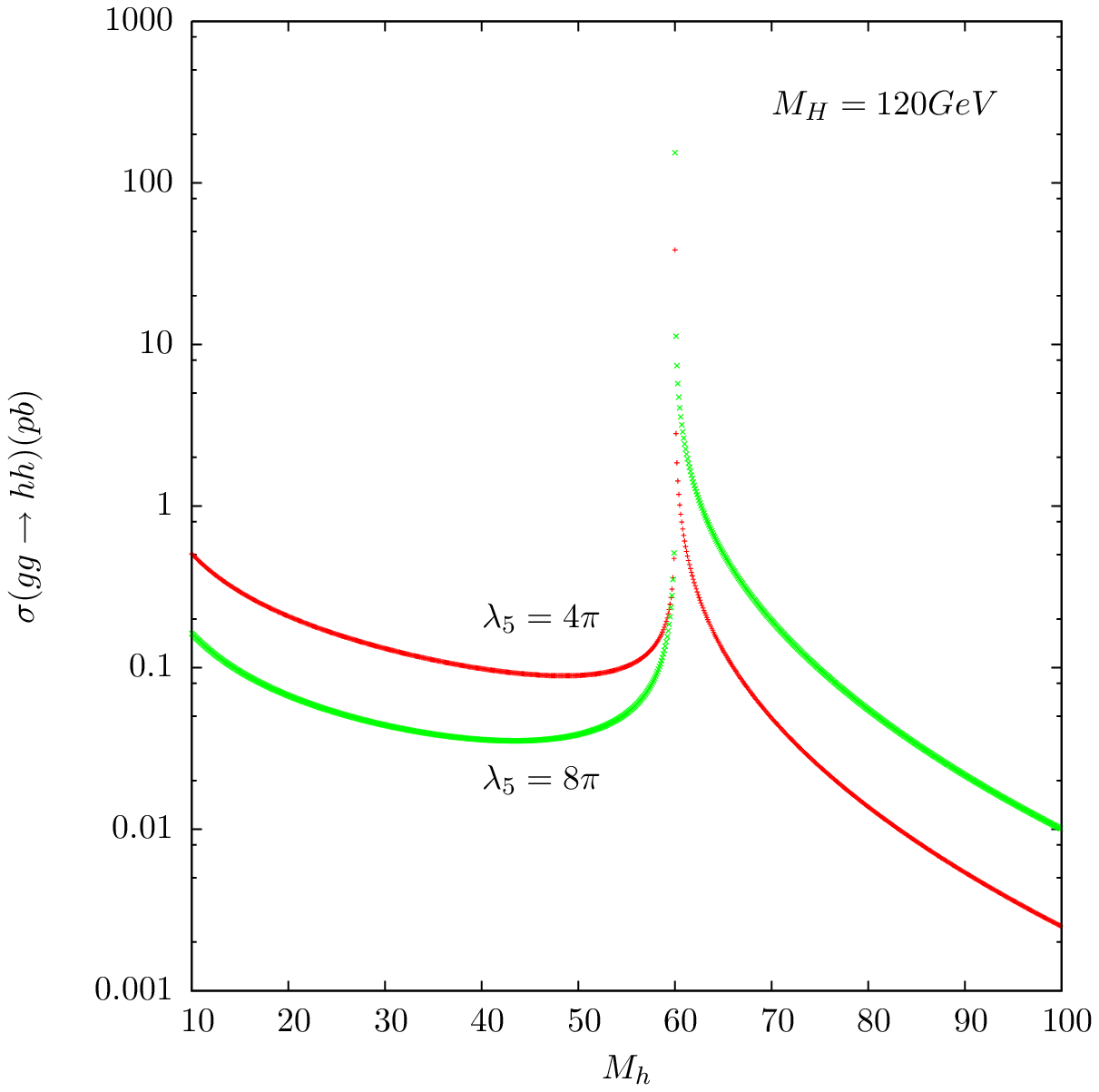}
\caption {$\sigma(gg \to hh)$ for the Tevatron in units of $pb$ as
a
  function of the
  fermiophobic Higgs mass with $\lambda_5 = 4
  \pi$ and $8 \pi$. On the left the mass of the heavier CP-even
  Higgs boson is twice the
  fermiophobic Higgs mass, i.e., on threshold, and on the right it is fixed (120 GeV).}
\label{plot1}
\end{figure}

\begin{figure}[t!]
\centering
\includegraphics[height=2.7in]{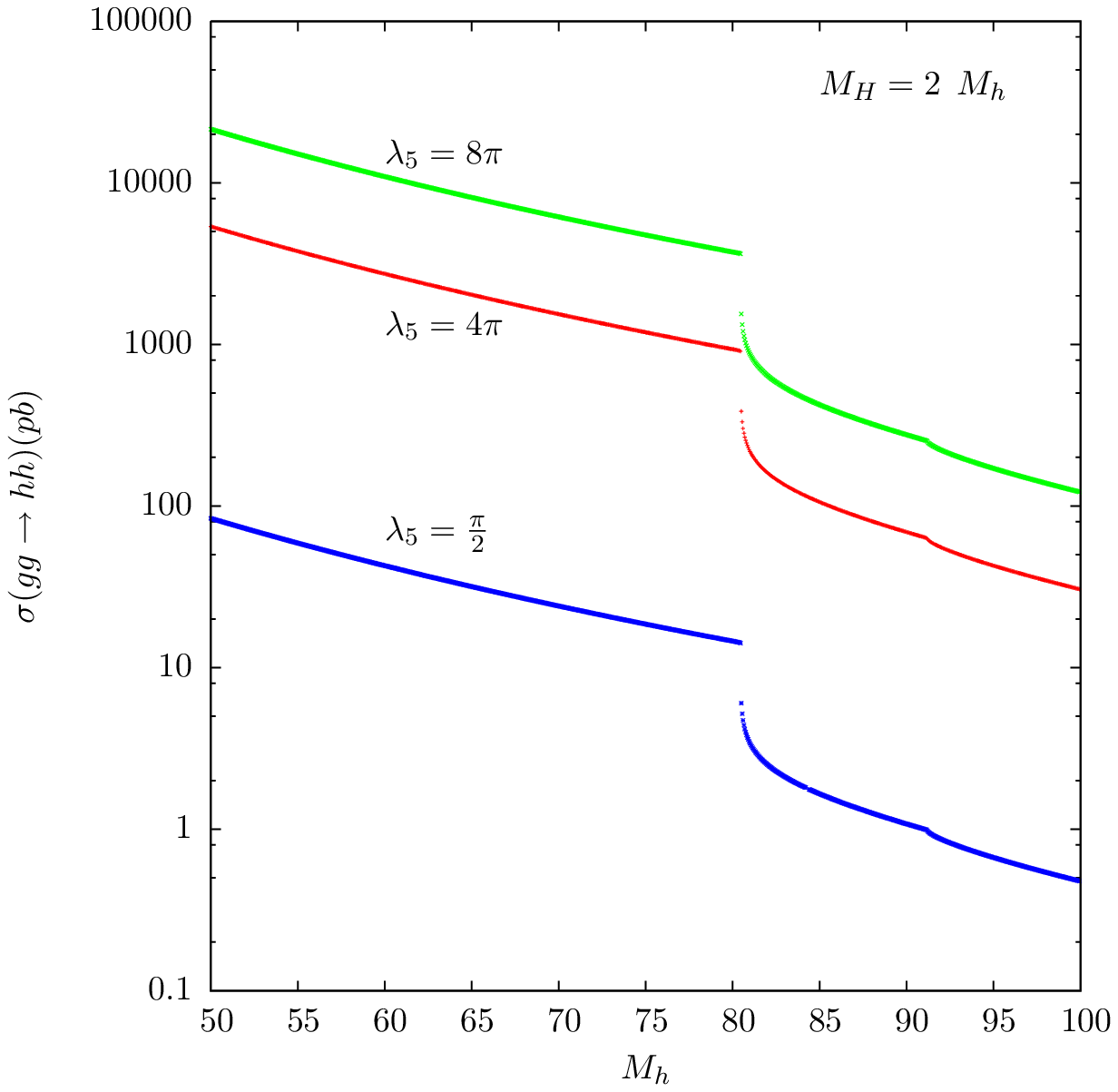}
\includegraphics[height=2.7in]{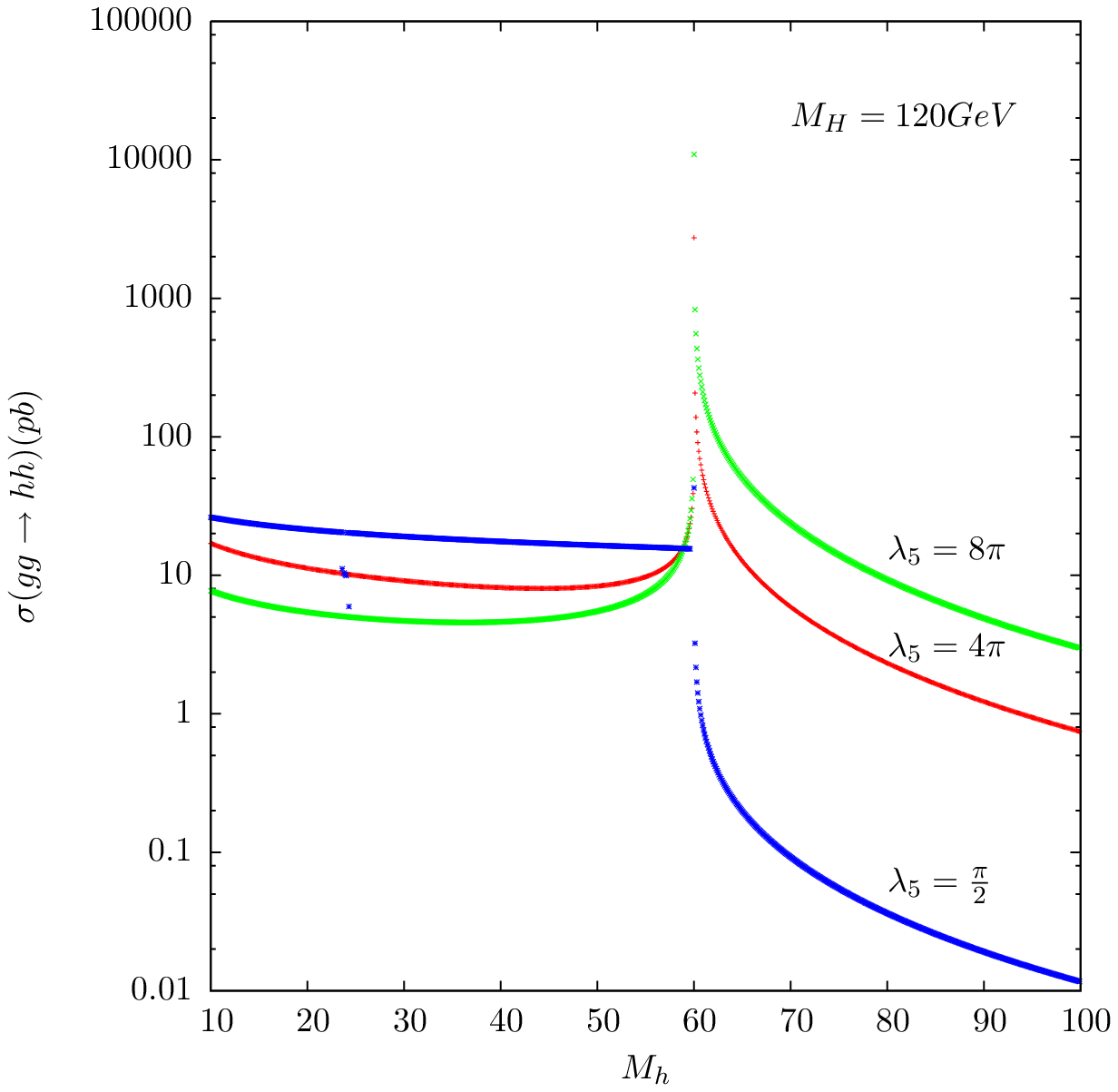}
\caption {$\sigma(gg \to hh)$ for the LHC in units of pb as a
  function of the
  fermiophobic Higgs mass with $\lambda_5 = \pi/2$ , $4
  \pi$ and $8 \pi$. On the left the mass of the heavier CP-even
  Higgs boson is twice the
  fermiophobic Higgs mass, i.e., on the threshold,and on the right it is fixed (120 GeV).}
\label{plot2}
\end{figure}

\noindent We first discuss $pp (\bar{p}) \to h h$ production. In
Fig.~\ref{plot1} (left) we illustrate the cross section of $p
\bar{p} \to h h$ as a function of the fermiophobic Higgs mass
$m_h$ for two representative values of $\lambda_5=4 \pi$ and $8
\pi$. The other heavy CP even mass is taken to be $m_H=2 m_h$ such
that the resonant channel $H\to h h$ is open. The other parameters
are: $m_A=m_{H\pm}=300$ GeV. On the $H$ resonance, the cross
section is enhanced and can reach a few hundreds of picobarn for a
very light fermiophobic Higgs $m_h=50$ GeV. Even for $m_h$ of the
order of 100 GeV, the cross section is still larger than 0.1 $pb$
which would give thousands of produced events for the planned
Tevatron luminosity of $10 fb^{-1}$. As can be seen from the
analytic expression of the coupling $H h h$, the large $\lambda_5$
is the large is the coupling $H h h$ and so is the $pp\to h h$
cross section. The observed kinks at $m_H=160$ and 182 GeV in the
left plot are due to the opening of the decay channels $H\to WW$
and $H\to ZZ$. On the right-hand side of Fig.~\ref{plot1} we show
the same cross section, the only difference being the $H$ mass
which is now fixed at 120 GeV. It is clear that even away from the
$H$ resonance a significant set of
 fermiophobic Higgs masses and $\lambda_5$ values can still be
probed at the Tevatron. The behavior with $\lambda_5$ changes at
threshold due to the $H$ width effect on the cross section. Above
threshold the $H \to hh$ channel is closed, the $H$ width is
 then very small. Therefore in that region the cross section is just
proportional to $\lambda_5^2$. Below threshold, the $H \to hh$
channel is open and the width in the $H$ propagator starts to play a
role which makes the dependence with $\lambda_5$ no longer
trivial.\\
 In Fig.~\ref{plot2} we show the same plots as in
Fig.~\ref{plot1} but for the LHC. As expected the plots are
re-scaled by more than one order of magnitude. For the values
shown, all masses between 10 and 100 GeV can be probed for most
values of $\lambda_5$. Even for $\lambda_5 = \pi/2$ the smallest
cross section value is of the order of 10 $fb$.
\begin{figure}[t!]
\centering
\includegraphics[height=2.7in]{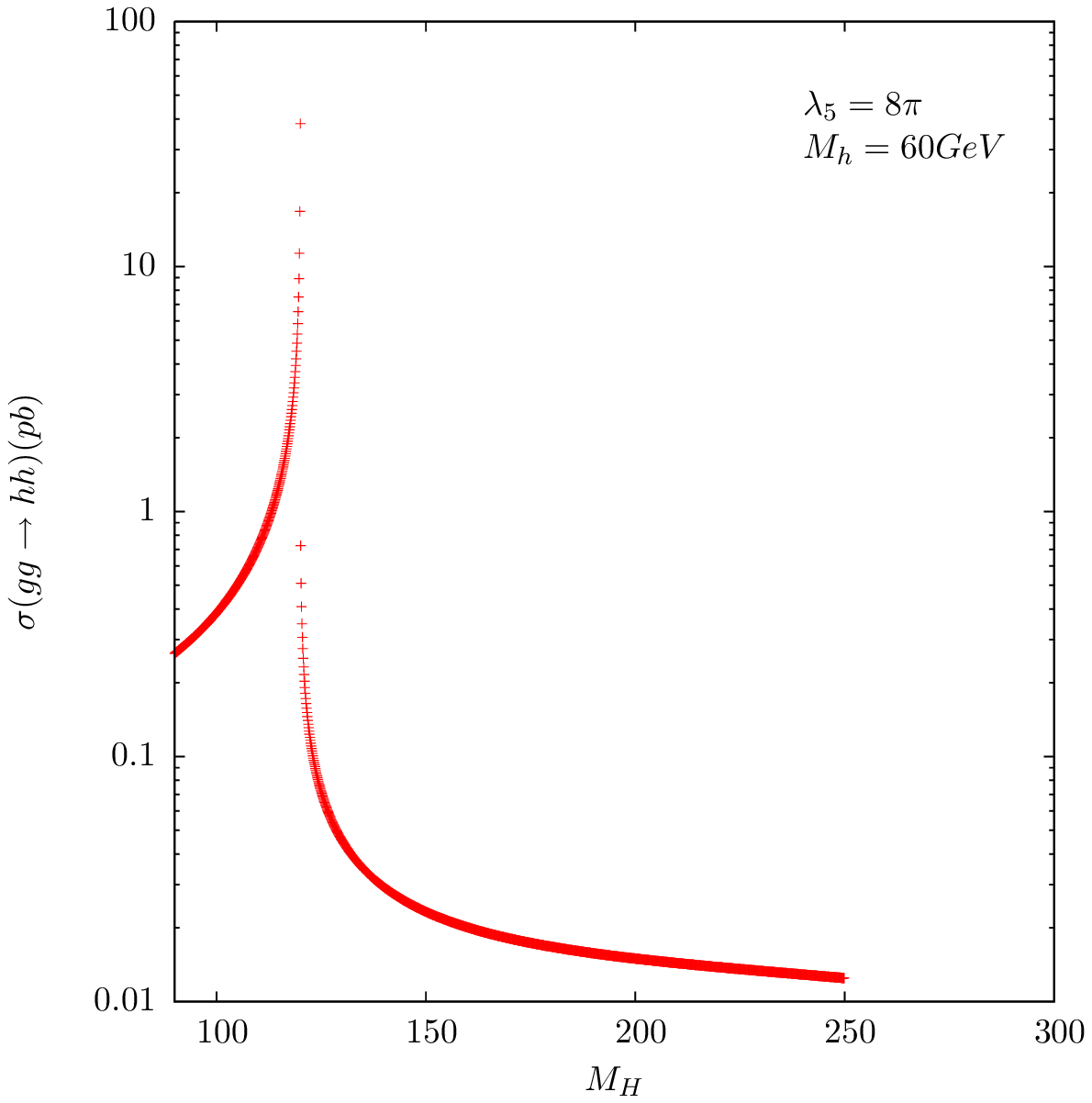}
\includegraphics[height=2.7in]{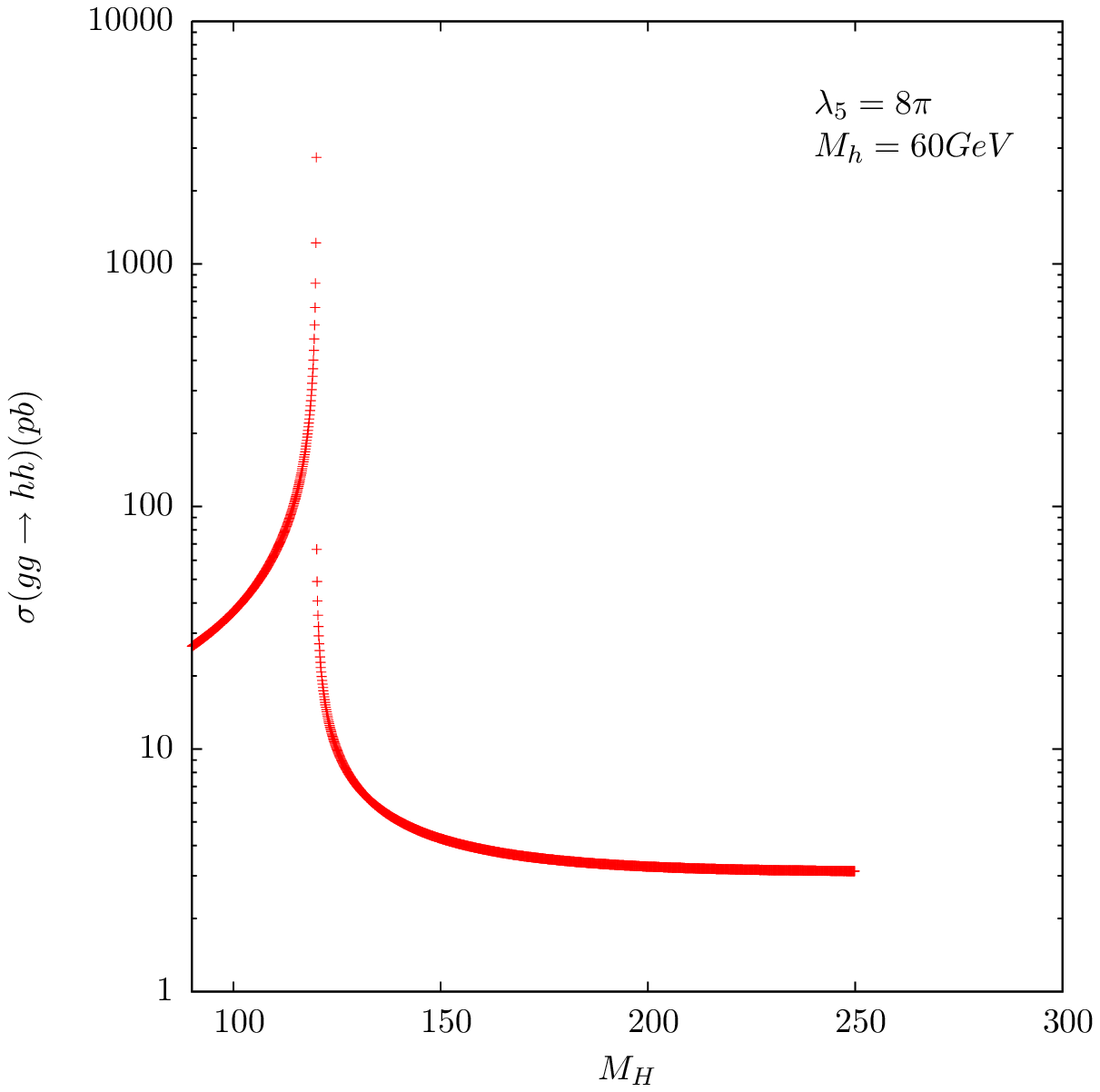}
\caption {$\sigma(gg \to hh)$ for the Tevatron (left) and for the
  LHC (right) in units of $pb$ as a
  function of the heavier CP-even Higgs mass with $\lambda_5 = 8 \pi$
  and a fermiophobic Higgs mass of 60 GeV.}
\label{plot3}
\end{figure}
%
\begin{figure}[t!]
\centering
\includegraphics[height=2.7in]{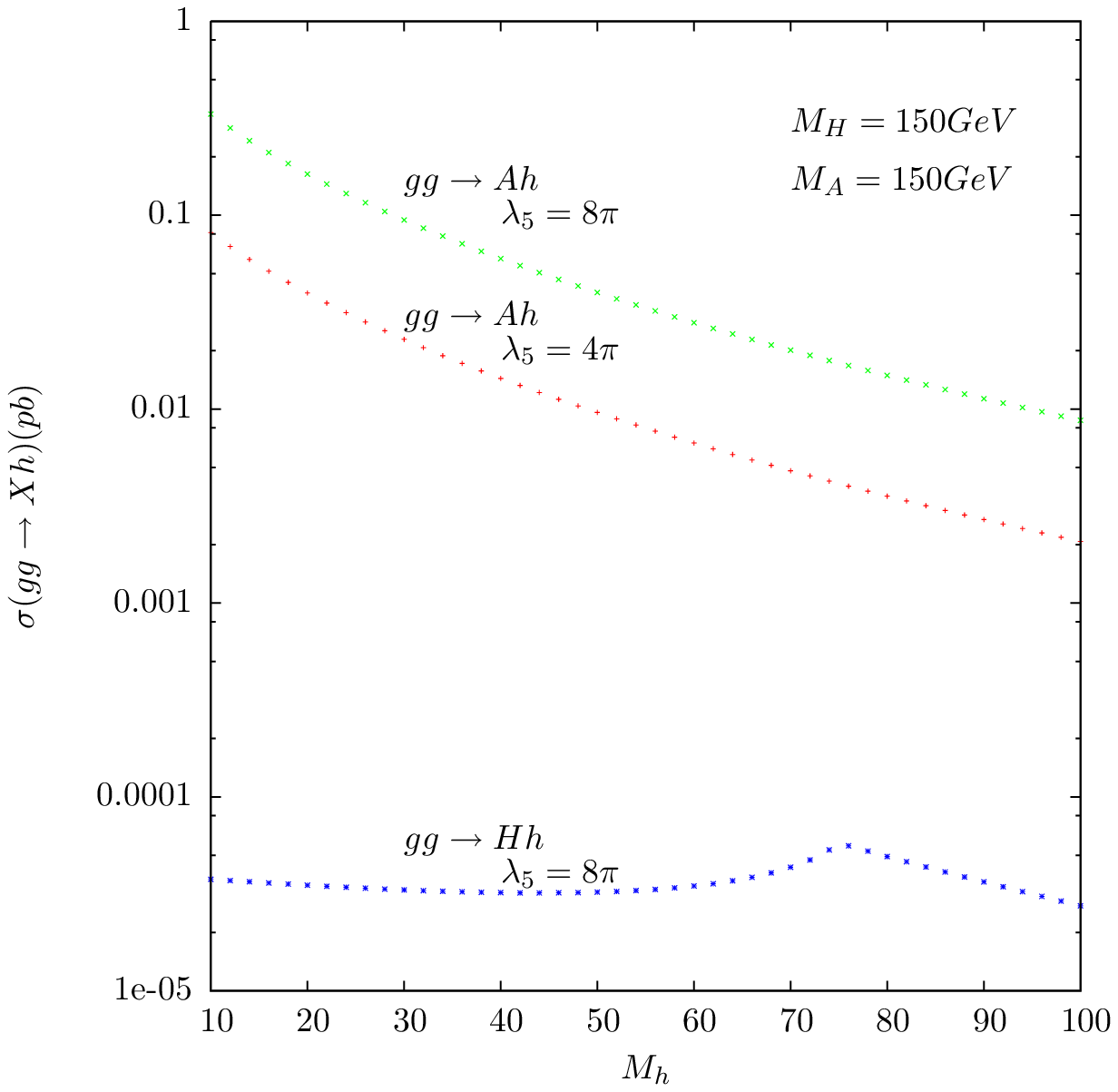}
\includegraphics[height=2.7in]{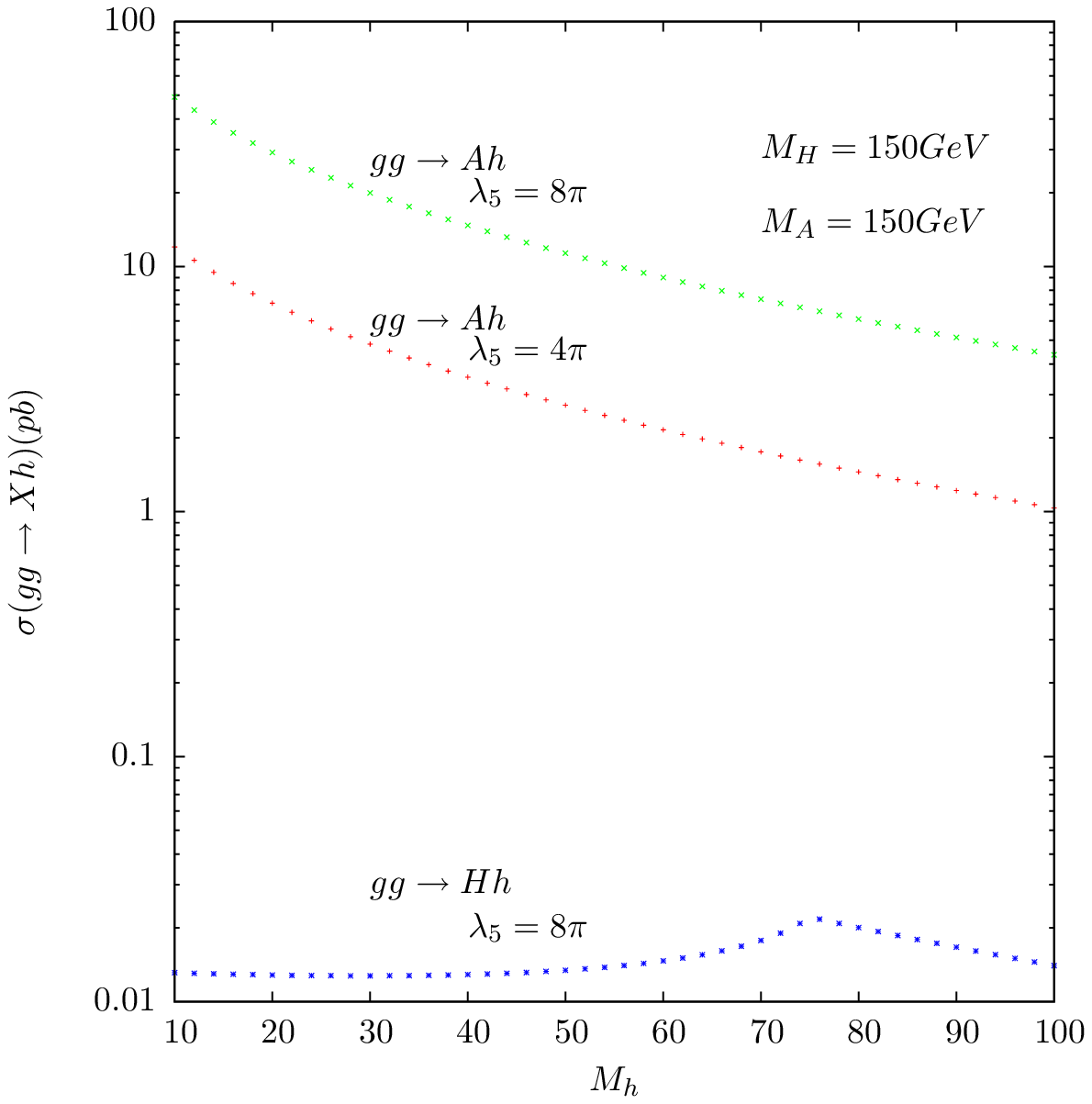}
\caption {$\sigma(gg \to hA)$ and $\sigma(gg \to hH)$ for the
  Tevatron (left) and for the
  LHC (right) in units of $pb$ as a
  function of the fermiophobic Higgs mass with
$\tan\beta = 10$, $\lambda_5 = 4 \pi, \, 8
  \pi$, $M_H = 150$ GeV and $M_A = 150$ GeV.}
\label{plot4}
\end{figure}

\noindent In Fig.~\ref{plot3} we show the cross section of $pp\to
hh$ as a function of the resonant Higgs mass $m_H$ for
$\lambda_5=8 \pi$ and  for a fermiophobic Higgs $m_h=60$ GeV, for
the Tevatron (left) and for the LHC (right). As one can see, away
from the resonance $m_H\approx 2 m_h$, the cross section is of the
order a few $fb$ for the Tevatron and a few $pb$ for the LHC. Once
we cross the resonance, one can see a spectacular pick for
$m_H=120$ GeV which is due to the opening of the decay channel
$H\to hh$. The peak is very sharp because the total width of $H$
is very narrow for $m_H\la 2 m_h$, less than $10^{-2}$ GeV. Once
the decay channel $H\to hh$ is open for $m_H\ga 2m_h$ the decay
width of $H$ increase suddenly to more than 100 GeV. This is
manifestly seen in the plot by a dramatic decrease of the cross
section from few hundred pb to $10^{-2}$ pb.

\noindent Finally we present the $pp (\bar{p}) \to Hh$ and $pp
(\bar{p}) \to Ah$ reactions. It is clear from both plots presented
in Fig.~\ref{plot4} that the cross section for $Hh$ production is
negligible for most of the parameter space. On the contrary, the
cross section for $hA$ production can be very large and still
within the Tevatron reach. The cross section $gg \to Ah$ can be
several orders of magnitude larger than the corresponding $q
\bar{q} \to Z^* \to Ah$ (see ref.~\cite{Akeroyd:2003bt}). We will
show in the next section that taking into account the
pseudo-scalar branching ratios, the decay $A\to Zh$ can be the
dominant one, and so the channel $gg \to Ah\to Zhh \to Z 4\gamma $
is still worth exploring at the Tevatron. Finally, we have checked
that the contribution from $q \bar{q} \to hh, \, Hh \, Ah$ was
negligible.

\section{Higgs signature}
\label{sec:sign}

\begin{figure}[t!]
\centering
\includegraphics[height=2.7in]{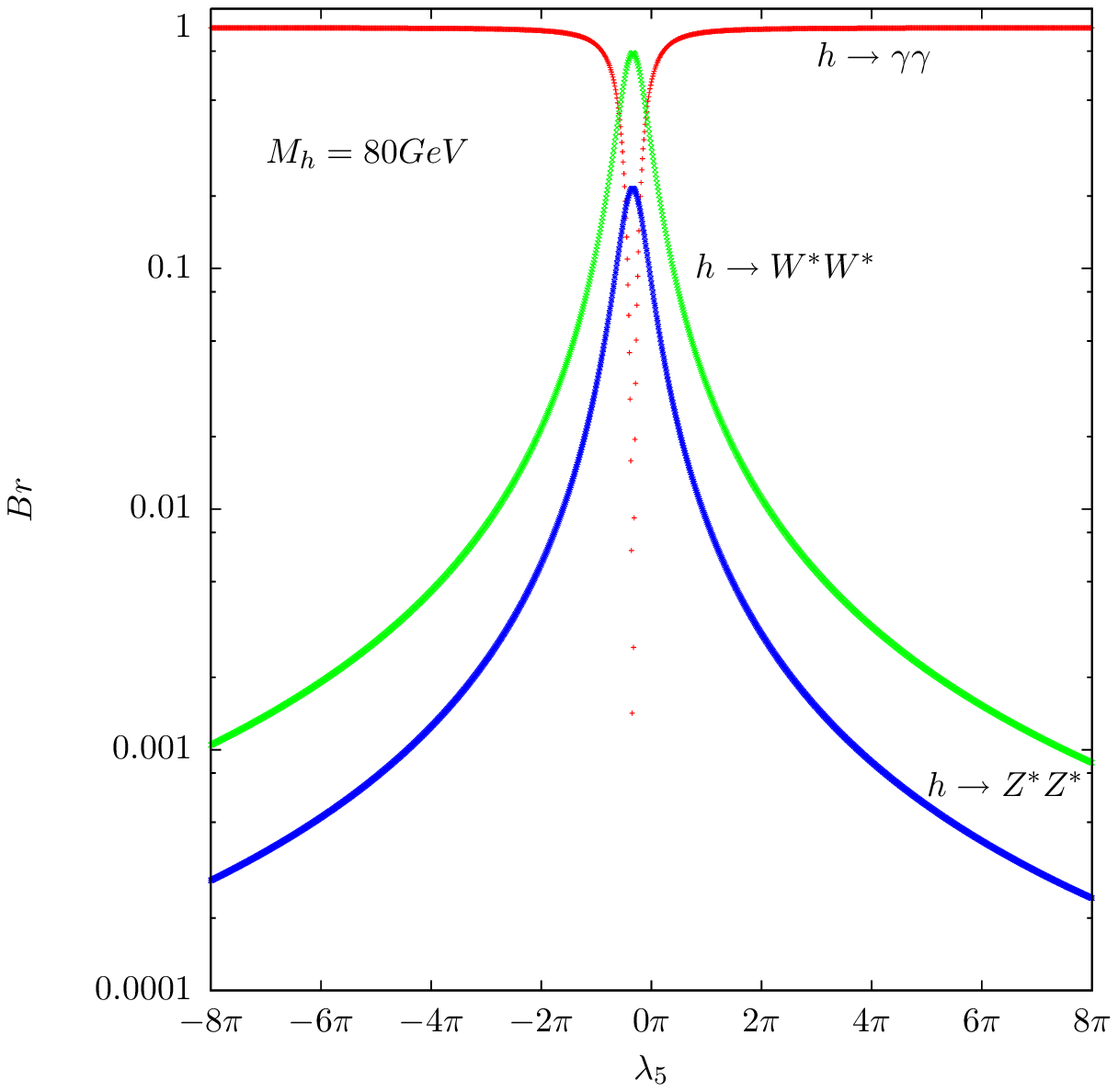}
\includegraphics[height=2.7in]{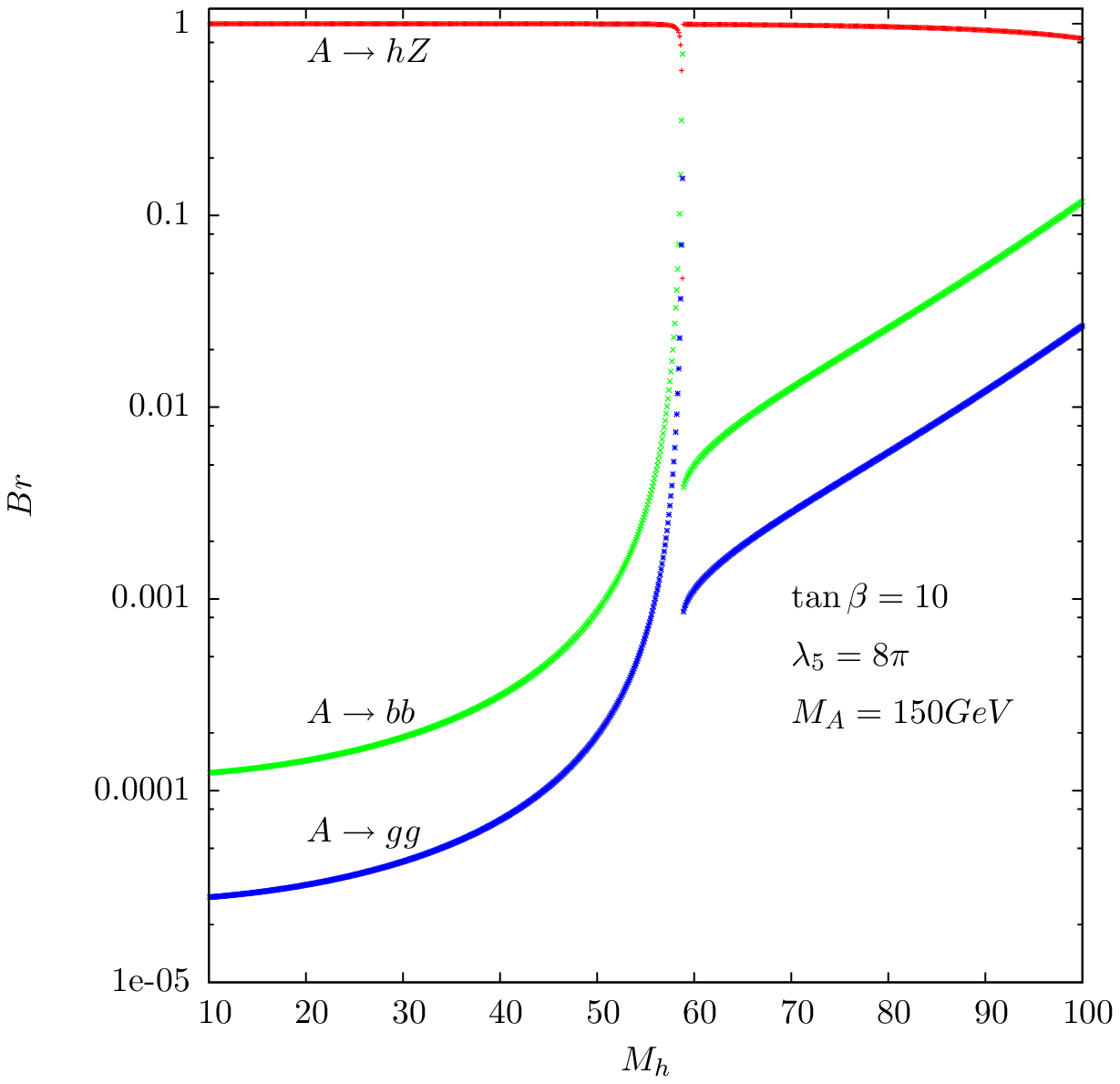}
\caption {Branching ratios of the fermiophobic Higgs (left) as a
function of $\lambda_5$ and of the CP-odd scalar as a function of
the fermiophobic Higgs mass. Both plots are for $\tan \beta = 10$,
$\lambda_5 = 8 \pi$, $M_H = M_A = M_{H^\pm} = 150$ GeV.}
\label{plot5}
\end{figure}

\noindent Having established that $gg \to hh$ and $gg \to Ah$ are
worth studying both at the Tevatron and at the LHC we now turn to
the experimental signatures for the fermiophobic Higgs and for the
pseudo-scalar in the parameter space under study. In the
fermiophobic limit there is a dramatic change in the fermiophobic
Higgs signatures. For smaller fermiophobic Higgs masses, the main
decay is to two photons (through W and charged Higgs loops) until
the WW channel starts to dominate. The crossing point of the
branching ratios depends on the parameters of the  scalar
potential which enters the game through the charged Higgs
contribution to $h\to \gamma\gamma$. Those parameters are mainly
the charged Higgs mass, $\lambda_5$, $\tan\beta$ and the
fermiophobic Higgs mass (see eq.(\ref{lhphm})). Recently a
detailed study of $h \to \gamma \gamma$ appeared
in~\cite{Akeroyd:2007yh}. Due to all experimental and theoretical
constraints, for a fermiophobic Higgs with mass between 10 and 100
GeV, $Br(h \to \gamma \gamma) \approx 100 \%$ except in a tiny
neighborhood of $\lambda_5 = 0$. In this neighborhood, all cross
sections are extremely small. Therefore there will always be a
tiny region around $\lambda_5=0$ that will not be probed with the
processes proposed here. For illustration, we show in
fig.~\ref{plot5} (left) the fermiophobic Higgs branching ratio as
a function of $\lambda_5$. We have checked that the larger $\tan
\beta$ is the higher the value of the $\gamma \gamma$ branching
ratio. $Br(h \to \gamma \gamma)$ decreases with the fermiophobic
Higgs mass, but for $m_h=100$ GeV the plot looks almost the same.
Regarding the pseudo-scalar decays, as one can see in
fig.~\ref{plot5} (right), one can have $Br(A \to h Z) \approx 100
\%$ if the decay $A\to H^\mp W^\pm$ is kinematically forbidden. We
have checked that changes in $\tan \beta$, $\lambda_5$ and the $A$
mass produce negligible changes in the branching ratio provided
the charged Higgs channel $A\to H^\mp W^\pm$ is kept closed.
Therefore, the 4 $\, \gamma$ final state is by far the dominant
one.

\section{Discussion and Conclusions}
\label{sec:con}

\noindent In this work we have shown that there are alternative
channels to search for fermiophobic Higgs with a multi-photon
signature. A vast region of the parameter space of the
fermiophobic THDM can be probed at the Tevatron and the analysis
can easily be extended for the LHC.
In the fermiophobic limit, the angle $\beta$ is already very
constrained. LEP~\cite{Rosca:2002me} has set a limit of $\tan
\beta > 10$ for almost all values of the fermiophobic Higgs masses
up to 100 GeV. For some masses the bound is even stronger. On the
other hand theoretical constraints tell us that these values
cannot be too high. This implies that $gg \to hh$ and $gg \to hA$
will be large while $gg \to h H$ will be negligible. Note however
that the $gg \to hh$ cross section does not depend on  $\tan\beta$
and that the $gg \to hA$ dependence on $\tan \beta$ for values
above 10 is negligible. Regarding the $\lambda_5$ dependence, we
have shown that there is a "tiny to small" region around
$\lambda_5=0$ that can not be probed. This is especially true for
$hh$ productions but $gg \to Ah$ decreases with $\lambda_5$ as
well.
\\
The process $p \bar{p} \rightarrow h H^{\pm} \rightarrow hh W^{\pm
\, *} \rightarrow 4 \gamma + X$ proposed in
\cite{Akeroyd:2003bt,Akeroyd:2005pr} and studied
in~\cite{D03gamma} is complementary to the processes we propose in
this work. We probe the region of small $H$ and/or $A$ masses
while \cite{D03gamma} probes the small charged Higgs mass region.
The advantage in our case is that our study is independent of
$\tan \beta$ while their process does not depend on $\lambda_5$.
Finally if all Higgs scalars besides the fermiophobic Higgs are
very heavy, only the two photon search can exclude a fermiophobic
Higgs.
\\
As expected and as it was shown in~\cite{D03gamma} the background
for a 3 or 4 photon final state is easy to control. In the case of
the 3 photon final state the main background contribution comes
from the direct tri-photon production (see~\cite{D03gamma} for
details).

\vspace{0.25cm} {\bf Acknowledgments:} Our thanks to Andrew
Akeroyd and Pedro Teixeira-Dias for a careful reading of the
manuscript and many interesting discussions. This work is
supported by Funda\c{c}\~ao para a Ci\^encia e Tecnologia under
contract POCI/FIS/59741/2004. R.S. is supported by FCT under
contract SFRH/BPD/23427/2005. R.G.J. is supported by FCT under
contract SFRH/BD/19781/2004. R.B. is supported by the National
Science Council of R.O.C under grant number: NSC96-2811-M-033-005.

\end{document}